\documentclass[secnumarabic, graphics,floatfix, nofootinbib,tightenlines,nobibnotes, aps, prl, 12pt]{revtex4-1}
\usepackage{graphicx}
\usepackage[english]{babel}
\usepackage[utf8]{inputenc}
\usepackage{amsmath,amssymb}
\usepackage{amsfonts}
\usepackage{multirow}

\begin{document}

\title{Event-plane dependent di-hadron correlations with harmonic $v_n$ subtraction in a hydrodynamic model}
\author{Wagner M. Castilho$^{1}$}
\author{Wei-Liang Qian$^{2,1}$}
\author{Yogiro Hama$^{3}$}
\author{Takeshi Kodama$^{4,5}$}

\affiliation{$^1$Faculdade de Engenharia de Guaratinguet\'a, Universidade Estadual Paulista, 12516-410, Guaratinguet\'a, SP, Brazil}
\affiliation{$^2$Escola de Engenharia de Lorena, Universidade de S\~ao Paulo, 12602-810, Lorena, SP, Brazil}
\affiliation{$^3$Instituto de F\'{\i}sica, Universidade de S\~ao Paulo, Caixa Postal 66318, 05389-970 S\~ao Paulo, SP, Brazil}
\affiliation{$^4$Instituto de F\'{\i}sica, Universidade Federal do Rio de Janeiro, Caixa Postal 68528, 21941-972 Rio de Janeiro, RJ, Brazil}
\affiliation{$^5$Instituto de F\'{\i}sica, Universidade Federal Fluminense, 24210-346, Niter\'oi, RJ, Brazil}

\date{Dec. 17, 2017}

\begin{abstract}
In this work, a hydrodynamic study of the di-hadron azimuthal correlations for the Au+Au collisions at 200 GeV is carried out.
The correlations are evaluated using the ZYAM method for the centrality windows as well as the transverse momentum range in accordance with the existing data.
Event-plane dependence of the correlation is obtained after the subtraction of contributions from the most dominant harmonic coefficients.
In particular, the contribution from the triangular flow, $v_3$, is removed from the proper correlations following the procedure implemented by the STAR collaboration.
The resultant structure observed in the correlations was sometimes attributed to the mini-jet dynamics, but the present calculations show that a pure hydrodynamic model gives a reasonable agreement with the main feature of the published data.  
A brief discussion on the physical content of the present findings is presented.

\end{abstract}

\maketitle

\section{Introduction}

Measurements on the two-particle correlations in the relativistic heavy-ion collisions, expressed in terms of the pseudorapidity difference $\Delta \eta$ and the angular spacing $\Delta \phi$, were carried out by various experimental collaborations \cite{star-ridge2,star-ridge3,phenix-ridge4,phenix-ridge5,phobos-ridge6,cms-ridge3,cms-ridge4} at both RHIC and LHC.
The shape of the two-particle correlations for different collision systems, at various transverse momentum range and its evolution as a function of trigger particle azimuthal angle have been both extensively studied.
They are understood to provide relevant information on the jets originating from small momentum transfer scatterings as well as the hot, dense medium created in the collisions \cite{qcd-corr-01,qcd-corr-02}.
The observed correlation yields are characterized by an enhancement on the near side around $\Delta \phi \approx 0$, known as the ``ridge", which possesses a long $\Delta \eta$ extension in the longitudinal direction.
Besides heavy-ion collisions, such ridge structures were also observed in pp \cite{cms-ridge7} and pA \cite{atlas-ridge-1,alice-ridge-1,cms-ridge9} collisions at LHC.
For pA collisions, it is found that the ridge yields vary with centrality.
The measured jet-like yields obtained by subtracting long-range pseudorapidity correlations observed in high-multiplicity events, on the other hand, are approximately constant for different centralities \cite{alice-ridge-2}.
This observed feature shows that the physics behind jet-like yield and ridge yield are indeed distinct, being the latter attributed to the collective flow of the system.
The correlation on the away side is found to be more significant in AA collisions than in pp and pA, and it presents a double-peak structure, usually called ``shoulders", which evolves continuously from the double peak for central to one peak for peripheral collisions \cite{phenix-ridge5,star-ridge3}.

In order to interpret the ``ridge" and ``shoulders" in AA collisions by a uniform picture, we proposed the so-called peripheral-tube model \cite{sph-corr-2,sph-corr-3,sph-corr-4,sph-corr-7,sph-vn-4}.
In this model, the phenomenon is attributed to the local (nonlinear) behavior of hydrodynamics.
The phenomenon can also be explained regarding the triangular flow, as usually done \cite{hydro-v3-1,hydro-v3-2}.
It is understood that hydrodynamical evolution transforms the spatial inhomogeneity of participating nucleons in the initial conditions into the momentum anisotropy of the observed hadrons \cite{hydro-v3-8,hydro-v3-5,hydro-v3-4}.
The triangular flow in the one-particle distribution function generates three peaks in the two-particle correlations: one peak on the near side at $\Delta\phi = 0$ and two others at $\Delta\phi = 2\pi/3$ and $\Delta\phi = 4\pi/3$ corresponding to the double peak on the away-side.
It is noted that studies by using the AMPT model were carried out \cite{hydro-v3-6,hydro-v3-7} which showed that the double peak disappears when the contributions due to the elliptic flow and triangular flow are subtracted.
The above finding seems to indicate that the triangular flow indeed plays an essential role in the observed structure on the away side.
The two-particle correlations are also investigated as a function of the trigger angle $\phi_{s}$, known as the event-plane dependence of the two-particle correlations \cite{star-plane1,star-plane2}.
It was found that the away side structure evolves from only one peak in the in-plane trigger direction with $\phi_{s}=0$ to double peaks at the out-of-plane trigger direction with $\phi_{s}=\pi/2$.
It is worth noting that the above data were extracted by mainly subtracting the background contributions of the elliptic flow.
The observed features of the data can be understood in terms of a hydrodynamic interpretation known as peripheral one-tube model \cite{sph-corr-ev-4} where geometric fluctuations are manifested as high-energy tubes randomly distributed in the initial conditions.
The model \cite{sph-corr-ev-6} is also able to explain the observed centrality dependence of the away side structure in the correlations \cite{phenix-ridge5}.

Based on analysis of the earlier STAR data~\cite{star-plane2}, Luzum pointed out~\cite{ph-corr-1} that the observed two-particle correlations are consistent with being entirely generated by the collective flow.
In the work, the author analyzed the elliptic flow coefficients $v_2^{(a)}, v_2^{(t,R)}$ and showed that the flow background used in ZYAM subtraction was probably underestimated, and the non-flow contribution of the second Fourier coefficient is likely insignificant.
However, it is not clear that the above reasoning also applies to the third Fourier coefficient, especially because no direct measurement of $v_3$ were made in STAR's analysis~\cite{star-plane2}.
This is because $V_{3\Delta}$ was found to be weakly dependent on $\phi_s$.
Even if $V_{3\Delta}$, the third Fourier coefficient of the proper correlations, and $v_3^{(a)} v_3^{(t,R)}$ have roughly the same dependence on $\phi_s$ provided they are not identical,  it is more likely that the subtraction was correct and the triangular non-flow signal is mostly independent of the trigger angle.
In other words, one does not know for sure that the non-flow contribution of the third Fourier coefficient is also insignificant.
Moreover, in our previous work, it is shown that~\cite{sph-corr-ev-4} $V_{3\Delta}$ is slightly dependent on $\phi_s$ while $v_3$ remains constant.
Such feature is consistent with the STAR data (see Fig.1 of ref.\cite{ph-corr-1}), and is understood as owing to an interplay between the multiplicity fluctuations and the background flow.
The above considerations, therefore, strengthen the role played by the triangular flow.
Recently, such a study was carried out by the STAR collaboration~\cite{star-plane3}.
In their work, in order to eliminate the contributions from the collective flow, the background includes those of the elliptic, triangular, and quadrangular flow.
It is somewhat surprising to find that resultant correlation yields still maintain the same feature on the away side: it shows one peak in the in-plane trigger direction and double peaks in the out-of-plane trigger direction.
In other words, after the subtraction of the contributions from the flow coefficients, some non-trivial structure remains.
As the STAR experiment claims, it may be attributed to the pathlength-dependent jet-quenching.
What we understand by ZYAM is that it is a method to single out the relevant signal in the resultant correlations, by subtraction of the background evaluated in terms of the {\it average flow} harmonics, from the proper correlations including the information on {\it event-by-event fluctuating flow}. 
Whether the method is good or not, we don't know. 
But the initial fluctuations should remain in the resultant correlations.
What can be drawn from the similarity shown in the two STAR papers~\cite{star-plane2,star-plane3} is the following. 
The contribution of the average triangular flow $\langle v_3 \rangle$ seems to be much smaller than $\langle v_2 \rangle$. 
Also, because of the different $|\Delta\eta|$ restrictions applied in the data analysis, the contribution from the jet should be either very small or insensitive to any $|\Delta\eta|$ cut.  
In this context, it is worthwhile to carry out an explicit calculation to verify whether the data presented in \cite{star-plane3} can be reproduced by a hydrodynamic approach.

The above discussions motivated the present study.
Here, we present results on event plane dependence of the di-hadron correlations by using NeXSPheRIO \cite{sph-review-1}, an ideal hydrodynamic model, with event-by-event fluctuating initial conditions.
Since all the parameters of the model are fixed by reproducing the observed multiplicity yields \cite{sph-eos-2,sph-eos-3}, there is no free parameter in the present calculation.
We then implement the same flow background due to the elliptic, triangular and quadrangular flows as the STAR collaboration did, and compare the calculated correlations with the data \cite{star-plane3}.
As a confirmation, we also evaluate and present the correlations by a subtraction only of the elliptic and quadrangular flows, in comparison with the data \cite{star-plane2}.
The rest of the paper is organized as follows.
In Section II, we show the results of our hydrodynamical simulations together with discussions, and the concluding remarks are given in Section III.

\section{Event plane dependence of di-hadron correlations}

The event-plane dependence analysis evaluates the correlated di-hadron pairs as a function of the azimuthal angle difference $\Delta \phi$ at different trigger angles $\phi_{s}=| \phi_{trigger}-\Psi_{2} |$, which is measured with respect to the event plane of the second harmonic coefficient.
In accordance with the experimental data \cite{star-plane3}, we study the collisions in the 20 - 60\% centrality window.
The transverse momentum range of the trigger particles is chosen to be $3 \le p_{T}^{t} \le 4\,$GeV, and that of the associated particles is $1 \le p_{T}^{a} \le 2\,$GeV.
To accommodate the experimental setup of the STAR collaboration, the calculations are carried out in the pseudorapidity interval $| \eta |< 1$.
Then the results are divided into six equally sized slices in the azimuthal angle of the trigger particles.
To calculate the event plane, we make use of all charged particles within the transverse momentum range $p_{T} < 2\,$Gev/c.
By carrying out the modified reaction-plane (MRP) method \cite{star-ridge6} as employed by the STAR collaboration, particle pairs with $| \Delta\eta | <$ 0.5 are excluded from the construction of the event planes.
Correlated pairs with $| \Delta \eta | < 0.7$ between the trigger particles and the associated particles are also excluded from the analysis because the original purpose was to minimize the near-side jet contribution \cite{star-plane3}.
The ZYAM method is then used to construct the correlation pattern due to the anisotropic flow.
The primary flow correlated background is from the elliptic flow, caused by the average almond shape of the superposition region of the initial energy distribution, the triangular flow, caused by the fluctuations of the initial conditions that occur on an event-by-event basis \cite{hydro-v3-1}, as well as the quadrangular flow.
The above flow harmonics contribute to the two-particle correlations and are to be subtracted from the proper correlation pattern.
In \cite{star-plane3}, the flow correlated background from the elliptic, triangular as well as quadrangular flow are expressed as follows
\begin{equation}
\frac{dN}{d\Delta\phi} = B \left(1+2v_{2}^{a}v_{2}^{t}\cos2\Delta\phi+2v_{3}^{a}v_{3}^{t}\cos3\Delta\phi+2v_{4}^{a} \{\Psi_{2}\} v_{4}^{t}\{\Psi_{2}\}\cos4\Delta\phi \right)    \label{zya1}
\end{equation}
where $B$ is the background normalization, $v_{2}^{a}$, $v_{4}^{a}\{\Psi_{2}\}$ ($v_{2}^{t}$ and $v_{4}^{t}\{\Psi_{2}\}$) are the second and fourth harmonic coefficients of the associated (trigger) particles with respect to the event plane of the second harmonic $\Psi_{2}$, $v_{3}^{t}$ and $v_{3}^{a}$ are triangular flow of the trigger and the associated particles, calculated with respect to the event plane of the third harmonic $\Psi_{3}$.
The harmonic coefficients of the trigger particle are the average value obtained in the respective slice of the azimuthal angle $\phi_{s}$.

\begin{figure}[h!]
\begin{center}
\includegraphics[width=17cm]{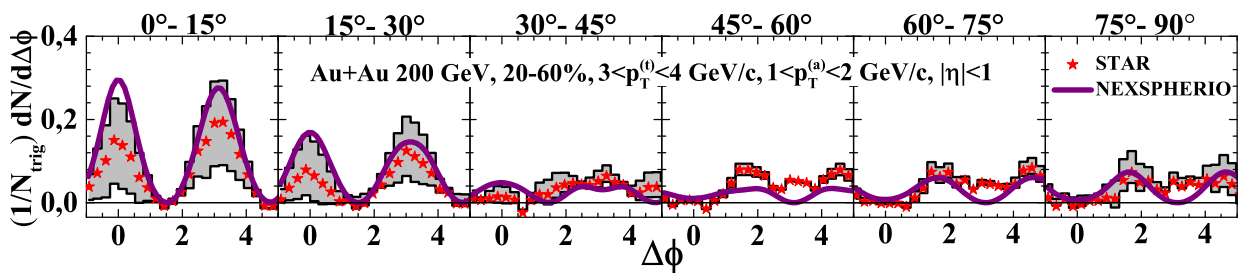}
\end{center}
\caption{The azimuthal di-hadron $\Delta \phi$ correlations for different values of $\phi_{s}$ = $| \phi_{t}-\psi_{2} |$.
The proper correlations are obtained by applying a cut on the pseudo-rapidity difference $| \Delta \eta | <$ 0.7 between the trigger and associated particles.
The resultant correlation is obtained by using ZYAM method, with $v_2$ and $v_3$ subtracted. 
The NeXSPheRIO results are shown by the solid purple curves, and the STAR data \cite{star-plane3} are represented by the red stars whereas the gray area between the solid lines indicates the uncertainties.}
\label{corep}
\end{figure}

In the calculations using NeXSPheRIO, the hydrodynamic simulations are carried out in smaller centrality bins of 10\% each.
The reason for such a division is to increase the statistics for more peripheral collisions, as we generate a total of 1300 events in the 20 - 30\% centrality window, 2400 events in the 30 - 40\% centrality windows, 4400 events in the of 40 - 50\% and 4600 events in the 50 - 60\% centrality windows.
To further increase the statistics, for each event with given initial conditions, the Monte Carlo hadron generator is invoked 200 times.
The obtained correlations of individual bins are then averaged to obtain the desired proper correlation.
To evaluate the background correlation, the harmonic coefficients are obtained by the event-plane method \cite{event-plane-method-1,event-plane-method-2,event-plane-method-3}.
Subsequently, the ZYAM method is made use of and is implemented according to Eq.(\ref{zya1}).
The resultant correlations are shown in Fig.\ref{corep}.
In the plots, the solid purple curves are those obtained by NeXSPheRIO and the data are represented by the red stars whereas the gray areas between the solid lines indicate the uncertainties.
Notice that, even though the contribution of the triangular flow is explicitly subtracted from the proper correlation as shown in (\ref{zya1}), the resultant correlation is still featured by one peak in the away-side for the in-plane direction, with its maximum at $\Delta \phi \approx \pi$, which evolves to double peaks for the out-of-plane direction.
Remark that similar feature had been reported by STAR collaboration in \cite{star-plane1,star-plane2} where only the elliptic and quadrangular flows were subtracted.

\begin{figure}[h!]
\begin{center}
\includegraphics[width=17cm]{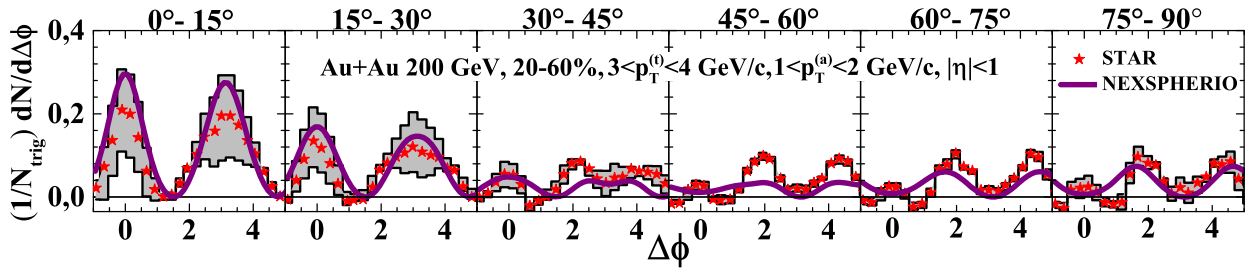}
\end{center}
\caption{The azimuthal di-hadron $\Delta \phi$ correlations for different values of $\phi_{s}= | \phi_{t}-\psi_{2} |$.
The correlations are obtained by using ZYAM method, with subtraction of $v_2$  
and $v_4$, but not $v_3$. The results are shown in solid purple curves.  The STAR data \cite{star-plane2} are represented by the red stars whereas the gray area between the solid lines indicates the uncertainties.}
\label{corm}
\end{figure}

As a comparison, we proceed to evaluate the two-particle correlation with the same specifications of \cite{star-plane2}.
We note that the experimental configurations, such as the pseudo-rapidity cut of $|\Delta\eta| > 0.7$ for the counting of particle pairs as well as the momentum ranges in consideration, are exactly the same in this case, and the differences come from the pseudo-rapidity filters in the construction of event planes.
Therefore the same data sets generated previously by the hydrodynamic simulations are used, while the event plane $\Psi_2$ is evaluated by taking into consideration all the particles while implementing a cut in pseudo-rapidity $|\Delta\eta| < 0.35$.
Both $v_2$ and $v_4$ are calculated with respected to the event plane $\Psi_2$.
Now, the correlated flow backgound is estimated as
\begin{equation}
\frac{dN}{d\Delta\phi} = B \left(1+2v_{2}^{a}v_{2}^{t}\cos2\Delta\phi+2v_{4}^{a} \{\Psi_{2}\} v_{4}^{t}\{\Psi_{2}\}\cos4\Delta\phi \right)    \label{zya2}
\end{equation}
The resulting two-particle correlations are shown in Fig.(\ref{corm}), compared with the STAR data from Ref.\cite{star-plane2}.
It is found that the double peak observed on the away side in the out-of-plane direction is also reasonably reproduced by the hydrodynamic calculations, which is consistent with our previous results \cite{sph-corr-ev-4} obtained by using the cumulant method.
It is somewhat surprising to find that the subtracted two-particle correlations in Fig.\ref{corep} and \ref{corm} are {\it not} very different.
This is because the triangular flow is understood to play a significant role in the explanation of the double-peak on the away side. Consequently, it was somehow expected that the subtraction of $v_3$ might lead to the disappearance of the same on the away side.
In order to quantitatively understand the contribution of $v_3$ in our hydrodynamical results, we present in Tables \ref{tb1} and \ref{tb2} the calculated values of flow harmonics used in the ZYAM subtraction to produce Fig.\ref{corep}.

\begin{table}[hbt]
\caption{The calculated flow harmonics $v_{n}$ of trigger particles for transverse momentum range of $3 < p_{T}^{t} < 4$ GeV/c for different azimuthal angles and centrality windows.}
\medskip
\begin{tabular}{c|c|c|c|c|c|c}
\hline \hline
\multicolumn{7}{c}{20 - 30\%} \\
\hline\hline

$\phi_{s} \equiv | \phi_{t}-\psi_{2} |$ & $0-\pi/12$ & $\pi/12-\pi/6$ & $\pi/6-\pi/4$ & $\pi/4-\pi/3$ & $\pi/3-5\pi/12$ & $5\pi/12-\pi/2$\\
\hline
$v_{2}$ & $0.1899$ & $0.1222$ & $0.0345$ & $-0.0235$ & $-0.0525$ & $-0.0649$\\
$v_{3}$ & $0.0304$ & $0.0065$ & $0.0065$ & $0.0314$ &$0.0288$& $0.0059$ \\
$v_{4}\{\Psi_{2}\}$& $0.1649$ & $0.0036$ & $-0.1043$ & $-0.0804$ & $-0.0011$ & $-0.0559$ \\
\hline
\hline \hline
\multicolumn{7}{c}{30 - 40\%} \\
\hline
$v_{2}$ & $0.2030$ & $0.1239$ & $0.0344$ & $-0.0226$ & $-0.0488$ & $-0.0594$\\
$v_{3}$ & $0.0311$ & $0.0073$ & $0.0066$ & $0.0298$ & $0.0297$ & $0.0068$ \\
$v_{4}\{\Psi_{2}\}$& $0.1767$ & $0.0038$ & $-0.1039$ & $-0.0779$ & $-0.0016$ & $-0.0513$ \\
\hline \hline
\multicolumn{7}{c}{40 - 50\%} \\
\hline
$v_{2}$ & $0.2038$ & $0.1277$ & $0.0345$ & $-0.0219$ & $-0.0497$ & $-0.0616$\\
$v_{3}$ & $0.0323$ & $0.0075$ & $0.0063$ & $0.0325$ & $0.0306$ & $0.0063$ \\
$v_{4}\{\Psi_{2}\}$& $0.1773$ & $0.0054$ & $-0.1028$ & $-0.0774$ & $-0.0008$ & $-0.0532$ \\
\hline \hline
\multicolumn{7}{c}{50 - 60\%} \\
\hline
$v_{2}$ & $0.2035$ & $0.1260$ & $0.0368$ & $-0.0236$ & $-0.0505$ & $-0.0664$\\
$v_{3}$ & $0.0313$ & $0.0052$ & $0.0069$ & $0.0295$ &$0.0282$& $0.0079$ \\
$v_{4}\{\Psi_{2}\}$& $0.1767$ & $0.0028$ & $-0.1075$ & $-0.0806$ & $-0.0014$ & $-0.0574$ \\
\hline

\end{tabular}
\label{tb1}
\end{table}

\begin{table}[hbt]
\begin{center}
\caption{The calculated flow harmonics $v_{n}$ of associated particles for transverse momentum range of $1 < p_{T}^{a} < 2$ GeV/c  for different azimuthal angles and centrality windows. }
\medskip
\begin{tabular}{c|c|c|c|c}
\hline \hline

 & 20 - 30\% & 30 - 40\% & 40 - 50\% & 50 - 60\% \\
\hline
$v_{2}$ & $0.1028$ & $0.1146$ & $0.1199$ & $0.1189$ \\
$v_{3}$ & $0.0399$ & $0.0421$ & $0.0445$ & $0.0449$  \\
$v_{4}\{\Psi_{2}\}$& $0.0106$ & $0.0138$ & $0.0144$ & $0.0135$ \\
\hline

\end{tabular}
\label{tb2}
\end{center}
\end{table}

We note that the data analysis carried out in this work is not for the most central window, and therefore, the average values of $v_3$ for the associated particles are smaller in comparison with those of $v_2$, as shown in Table \ref{tb2}.
This is also the case for the trigger particles at the in-plane directions, as can be readily verified in the first few columns of Table \ref{tb1}.
One also observes that the magnitude of $v_4\{\Psi\}$ for the associated particles follows the same trend of that of $v_2$.
However, since its size is much smaller than that of $v_2$, it does not play a significant role in the resultant correlations.
On the other hand, for the trigger particles, as one goes to the out-of-plane directions, the elliptic flow $v_2^t$ starts to decrease and eventually becomes negative, as confirmed by the three rightmost columns in Table \ref{tb1}.
This is expected, since $v_2^t$ is evaluated with respect to the event plane $\Psi_2$ of the associated particles, so this is implied by definition.
Owing to the change of the sign of $v_2$, the contribution of $v_3$ in the background Eq.(\ref{zya1}) is relatively small in the in-plane direction in comparison to that in the out-of-plane direction.
On the other hand, for the calculated correlations, the appearance of the double peak in the out-of-plane direction indicated that the observed modulation of the resultant correlation is dominated by the second order harmonic coefficient $V_2\equiv v_{2}^{a}v_{2}^{t}$ in the in-plane direction and the remnant of $V_3\equiv v_{3}^{a}v_{3}^{t}$ in the out-of-plane direction.
By Eq.(\ref{zya1}), this is consistent with the previous observations on the magnitudes of $v_2$ and $v_3$.

In Ref.\cite{star-plane3}, it is discussed that the trends of the away-side correlation might underscore the importance of path-length-dependent jet-medium interactions.
As the present hydrodynamical simulations are able to reproduce observed feature of two-particle correlation, it strongly indicates that the observed correlations are likely to be a collective-flow effect of the system.
Moreover, it seems that the ZYAM procedure, devised to essentially subtract the contribution of collective flow from the proper two-particle correlation, has somehow failed in its purpose.
In particular, it is found that the subtraction of $v_3$ does not affect the essential feature of the resultant correlations, namely, the relative magnitudes between $V_2$ and $V_3$.
To us, this might be related to the event-by-event fluctuating initial conditions and their impact on flow harmonics.
This is because the subtracted $v_n$ in Eq.(\ref{zya1}) is, in fact, the event average value, $\langle v_n \rangle$.
Due to the event-by-event fluctuations, the event average value of a product of harmonic coefficients can be significantly different from the product of the corresponding average values.
In other words, not only the magnitude of the triangular flow, $v_3$, is understood to be related to the event-by-event fluctuations, its fluctuations might also play a non-trivial role in the particle correlations.
Moreover, the event planes between different harmonics might be correlated for a given event but uncorrelated among the various events, which further complicates the problem.
As a result, the average of the Fourier expansion in azimuthal angle cannot be simply approximated by a Fourier expansion in terms of the products of average harmonic coefficients, even rescaled by the ZYAM scheme.
The present calculations employing NeXSPheRIO give reasonable results for two different sets of data obtained by different procedures.
It implies that the fluctuating initial conditions generated by NeXuS are mostly realistic.
We also note that this topic is closely related to the correlation between different flow harmonics, as recently explored by several authors \cite{atlas-vn3,atlas-vn5,alice-vn5}.

It is noted that in the above calculations, the event-plane method is employed for the background subtraction, as a part of our strategy to closely follow the same steps in the data analysis.
However, the event-plane method carried out in a hydrodynamic study might not be equivalent to that in experiment analysis.
According to the calculations by the Monte Carlo Glauber model~\cite{phobos-mcglauber-1}, the obtained $v_2$ depends on the resolution $R$ of the event-plane method.
As long as the resolution is high enough, the calculated result gives the desired mean value of $v_n$.
However, as the resolution decreases, the obtained result gradually approaches the RMS value of corresponding harmonic coefficients.
Since in a hydrodynamic calculation, the resolution is typically better than that of experimental measurements, the current calculations likely underestimate the flow background.
In practice, the resolution of a hydrodynamic approach can be controlled by the number of Monte Carlo simulations during the hadronization phase.
One may therefore roughly estimate the discrepancy owing to the resolution, by extrapolating the present results to the corresponding resolution of the data by varying the number of Monte Carlo simulations.
We have carried out the analysis, and it shows that the flow harmonics used in the flow background subtraction is around 10\% smaller than those employed in STAR analysis.
Since the magnitudes of flow harmonics shown in Tables \ref{tb1} and \ref{tb2} are small, the deviations of their contributions in Eqs.(\ref{zya1}) and (\ref{zya2}) are even less.
As a result, the values of $B$ in Eqs.(\ref{zya1}) and (\ref{zya2}) should be slightly smaller but mostly unchanged.
We have verified that the resultant correlation after the flow background subtraction will be slightly smaller, but the qualitative results shown in Fig.\ref{corep} and \ref{corm} remain unchanged.

\section{Concluding remarks}

As discussed above, we understand that ZYAM is a method aimed to single out relevant signal in the resultant correlations, by subtraction of the flow background evaluated in terms of the average harmonic coefficients, from the proper correlations including the event-by-event fluctuating flow.
To us, it seems possible that the initial fluctuations still present, or even become dominant in the resultant correlations.
The similarity presented in the two STAR papers~\cite{star-plane2,star-plane3} indicates that the flow is fluctuating and $\langle v_3 \rangle \ll \langle v_2 \rangle$.
Also, owing to the different $|\Delta\eta|$ restrictions applied in the data analysis, the jet contribution is likely to be very insignificant.
In this work, we show by explicit calculations that hydrodynamics, with fluctuating initial conditions, is able to reproduce the observed double-peak structure of two-particle correlation on the away-side for the out-of-plane triggers, even when the triangular flow is subtracted by using ZYAM method.
Therefore, the present study further strengthens the idea that observed correlations are mostly of hydrodynamic origin.

From a hydrodynamic viewpoint, it is understood that the role of $v_3$ is closely associated with the event-by-event geometrical fluctuations in the initial conditions.
The existence of two peaks, in the away side correlation, clearly shows the need of $v_3$.
Our results indicate that the observed event plane dependence of two-particle correlations, for the most part, can be reproduced by a hydrodynamic approach.
The physical mechanism behind the findings may be attributed to the correlation and/or fluctuation of flow harmonics, which is closely associated with the event-by-event fluctuations in the initial conditions.
This is because of the assumption of the background correlation module, Eq.(\ref{zya1}), assumes the picture of average flow with no correlation neither fluctuation of significant consequence.
Ongoing efforts on the event-plane correlations, symmetric cumulant  \cite{atlas-vn3,atlas-vn5,alice-vn5} may likely provide a better insight of the problem, and a clear criterion to distinguish between different mechanisms.
Some alternative methods may also be meaningful, as shown in our recent work concerning centrality dependence of di-hadron correlations~\cite{sph-corr-ev-6}.
There, the cumulant method gives quite a similar result as the ZYAM one.
We note that the 2+1 correlation \cite{star-corr-2plus1-1,star-corr-2plus1-2} employed by STAR Collaboration may also serve as a good tool to disentangle the jet signal.

On the other hand, the peripheral tube model \cite{sph-corr-2}, in which the interplay between random hot tubes in the initial conditions and collective background flow plays an important role, gives a unified description of the ``ridge'' structure, both for the near-side and the away-side ones.
As a result, in this interpretation, different flow harmonics of an individual event are naturally ``born" together, and consequently correlated, their appearance does not depend on any global structure of the initial condition.
From the viewpoint of the peripheral tube model, it is possible that the correlation between $v_2$ and $v_3$ owing to the flow deflected by high energy tube may provide an intuitive explanation of the observed data.
In particular, the lack of the near-side peak for the out-of-plane triggers, which would be produced by this two-peak structure in the single-particle distribution, might be studied analytically in this framework.
Recently, a Reaction Plane Fit method was proposed~\cite{zyam-rpf-1} and employed to estimate the correlation functions in the background dominated region on the near-side~\cite{zyam-rpf-2}.
The resulting correlation does not show the double peak on the away side, neither any dramatic shape modification as a function of centrality.
Therefore, the authors conclude that the Mach cone is an artifact of the background subtraction and the jets do not fully equilibrate with the medium.
These results further indicate that the effect of the jet in the di-hadron correlation is indeed a subtle subject. We plan to explore these topics further and look for a possible criterion to distinguish between different approaches in the near future.

\section*{Acknowledgments}
We are thankful for valuable discussions with Matthew Luzum and Jiangyong Jia.
We gratefully acknowledge the financial support from
Funda\c{c}\~ao de Amparo \`a Pesquisa do Estado de S\~ao Paulo (FAPESP),
Funda\c{c}\~ao de Amparo \`a Pesquisa do Estado do Rio de Janeiro (FAPERJ),
Conselho Nacional de Desenvolvimento Cient\'{\i}fico e Tecnol\'ogico (CNPq),
and Coordena\c{c}\~ao de Aperfei\c{c}oamento de Pessoal de N\'ivel Superior (CAPES).
A part of the work was developed under the project INCTFNA Proc. No. 464898/2014-5.
This research is also supported by the Center for Scientific Computing (NCC/GridUNESP) of the S\~ao Paulo State University (UNESP).

\bibliographystyle{h-physrev}
\bibliography{references_qian}{}

\begin{thebibliography}{10}

\bibitem{star-ridge2}
STAR Collaboration, B.~Abelev {\em et~al.},
\newblock Phys.Rev. {\bf C80}, 064912 (2009), arXiv:0909.0191.

\bibitem{star-ridge3}
STAR Collaboration, B.~Abelev {\em et~al.},
\newblock Phys.Rev.Lett. {\bf 102}, 052302 (2009), arXiv:0805.0622.

\bibitem{phenix-ridge4}
PHENIX Collaboration, A.~Adare {\em et~al.},
\newblock Phys.Rev. {\bf C77}, 011901 (2008), arXiv:0705.3238.

\bibitem{phenix-ridge5}
PHENIX Collaboration, A.~Adare {\em et~al.},
\newblock Phys.Rev. {\bf C78}, 014901 (2008), arXiv:0801.4545.

\bibitem{phobos-ridge6}
PHOBOS Collaboration, B.~Alver {\em et~al.},
\newblock Phys.Rev.Lett. {\bf 104}, 062301 (2010), arXiv:0903.2811.

\bibitem{cms-ridge3}
CMS Collaboration, S.~Chatrchyan {\em et~al.},
\newblock Eur.Phys.J. {\bf C72}, 2012 (2012), arXiv:1201.3158.

\bibitem{cms-ridge4}
CMS Collaboration, S.~Chatrchyan {\em et~al.},
\newblock JHEP {\bf 1107}, 076 (2011), arXiv:1105.2438.

\bibitem{qcd-corr-01}
E.~V. Shuryak,
\newblock Phys. Rev. {\bf C76}, 047901 (2007), 0706.3531.

\bibitem{qcd-corr-02}
C.~B. Chiu and R.~C. Hwa,
\newblock Phys. Rev. {\bf C79}, 034901 (2009), 0809.3018.

\bibitem{cms-ridge7}
CMS, V.~Khachatryan {\em et~al.},
\newblock JHEP {\bf 09}, 091 (2010), arXiv:1009.4122.

\bibitem{atlas-ridge-1}
ATLAS, G.~Aad {\em et~al.},
\newblock Phys. Rev. Lett. {\bf 110}, 182302 (2013), arXiv:1212.5198.

\bibitem{alice-ridge-1}
ALICE, B.~Abelev {\em et~al.},
\newblock Phys. Lett. {\bf B719}, 29 (2013), arXiv:1212.2001.

\bibitem{cms-ridge9}
CMS, S.~Chatrchyan {\em et~al.},
\newblock Phys. Lett. {\bf B718}, 795 (2013), arXiv:1210.5482.

\bibitem{alice-ridge-2}
ALICE, B.~B. Abelev {\em et~al.},
\newblock Phys. Lett. {\bf B741}, 38 (2015), arXiv:1406.5463.

\bibitem{sph-corr-2}
R.~Andrade, F.~Grassi, Y.~Hama, and W.-L. Qian,
\newblock J.Phys.G {\bf G37}, 094043 (2010), arXiv:0912.0703.

\bibitem{sph-corr-3}
Y.~Hama, R.~P.~G. Andrade, F.~Grassi, and W.-L. Qian,
\newblock Nonlin.Phenom.Complex Syst. {\bf 12}, 466 (2009), arXiv:0911.0811.

\bibitem{sph-corr-4}
R.~P.~G. Andrade, F.~Grassi, Y.~Hama, and W.-L. Qian,
\newblock Phys.Lett. {\bf B712}, 226 (2012), arXiv:1008.4612.

\bibitem{sph-corr-7}
Y.~Hama, R.~P. Andrade, F.~Grassi, J.~Noronha, and W.-L. Qian,
\newblock Acta Phys.Polon.Supp. {\bf 6}, 513 (2013), arXiv:1212.6554.

\bibitem{sph-vn-4}
W.-L. Qian {\em et~al.},
\newblock J.Phys.G {\bf G41}, 015103 (2014), arXiv:1305.4673.

\bibitem{hydro-v3-1}
B.~Alver and G.~Roland,
\newblock Phys.Rev. {\bf C81}, 054905 (2010), arXiv:1003.0194.

\bibitem{hydro-v3-2}
D.~Teaney and L.~Yan,
\newblock Phys.Rev. {\bf C83}, 064904 (2011), arXiv:1010.1876.

\bibitem{hydro-v3-8}
H.~Petersen, G.-Y. Qin, S.~A. Bass, and B.~Muller,
\newblock Phys. Rev. {\bf C82}, 041901 (2010), arXiv:1008.0625.

\bibitem{hydro-v3-5}
B.~Schenke, S.~Jeon, and C.~Gale,
\newblock Phys.Rev.Lett. {\bf 106}, 042301 (2011), arXiv:1009.3244.

\bibitem{hydro-v3-4}
G.-Y. Qin, H.~Petersen, S.~A. Bass, and B.~Muller,
\newblock Phys.Rev. {\bf C82}, 064903 (2010), arXiv:1009.1847.

\bibitem{hydro-v3-6}
J.~Xu and C.~M. Ko,
\newblock Phys.Rev. {\bf C83}, 021903 (2011), arXiv:1011.3750.

\bibitem{hydro-v3-7}
G.-L. Ma and X.-N. Wang,
\newblock Phys.Rev.Lett. {\bf 106}, 162301 (2011), arXiv:1011.5249.

\bibitem{star-plane1}
STAR Collaboration, A.~Feng,
\newblock J.Phys.G {\bf G35}, 104082 (2008), arXiv:0807.4606.

\bibitem{star-plane2}
STAR Collaboration, H.~Agakishiev {\em et~al.},
\newblock (2010), arXiv:1010.0690.

\bibitem{sph-corr-ev-4}
W.-L. Qian, R.~Andrade, F.~Gardim, F.~Grassi, and Y.~Hama,
\newblock Phys.Rev. {\bf C87}, 014904 (2013), arXiv:1207.6415.

\bibitem{sph-corr-ev-6}
W.~M. Castilho, W.-L. Qian, F.~G. Gardim, Y.~Hama, and T.~Kodama,
\newblock Phys.Rev. {\bf C95}, 064908 (2017), arXiv:1610.04108.

\bibitem{ph-corr-1}
M.~Luzum,
\newblock Phys.Lett. {\bf B696}, 499 (2011), arXiv:1011.5773.

\bibitem{star-plane3}
STAR, H.~Agakishiev {\em et~al.},
\newblock Phys. Rev. {\bf C89}, 041901 (2014), arXiv:1404.1070.

\bibitem{sph-review-1}
Y.~Hama, T.~Kodama, and O.~Socolowski~Jr.,
\newblock Braz.J.Phys. {\bf 35}, 24 (2005), arXiv:hep-ph/0407264.

\bibitem{sph-eos-2}
W.-L. Qian {\em et~al.},
\newblock Braz.J.Phys. {\bf 37}, 767 (2007), arXiv:nucl-th/0612061.

\bibitem{sph-eos-3}
D.~M. Dudek {\em et~al.},
\newblock (2014), arXiv:1409.0278.

\bibitem{star-ridge6}
STAR, J.~Adams {\em et~al.},
\newblock Phys. Rev. {\bf C72}, 014904 (2005), arXiv: nucl-ex/0409033.

\bibitem{event-plane-method-1}
S.~Voloshin and Y.~Zhang,
\newblock Z. Phys. {\bf C70}, 665 (1996), arXiv:hep-ph/9407282.

\bibitem{event-plane-method-2}
A.~M. Poskanzer and S.~A. Voloshin,
\newblock Phys. Rev. {\bf C58}, 1671 (1998), arXiv:nucl-ex/9805001.

\bibitem{event-plane-method-3}
S.~A. Voloshin, A.~M. Poskanzer, and R.~Snellings,
\newblock (2008), arXiv:0809.2949.

\bibitem{atlas-vn3}
ATLAS, G.~Aad {\em et~al.},
\newblock Phys. Rev. {\bf C90}, 024905 (2014), arXiv:1403.0489.

\bibitem{atlas-vn5}
J.~Jia,
\newblock J. Phys. {\bf G41}, 124003 (2014), arXiv:1407.6057.

\bibitem{alice-vn5}
ALICE, J.~Adam {\em et~al.},
\newblock Phys. Rev. Lett. {\bf 117}, 182301 (2016), arXiv:1604.07663.

\bibitem{phobos-mcglauber-1}
B.~Alver {\em et~al.},
\newblock Phys.Rev. {\bf C77}, 014906 (2008), arXiv:0711.3724.

\bibitem{star-corr-2plus1-1}
STAR, H.~Pei,
\newblock J. Phys. Conf. Ser. {\bf 316}, 012016 (2011).

\bibitem{star-corr-2plus1-2}
STAR, K.~Kauder,
\newblock Nucl. Phys. {\bf A830}, 685C (2009), arXiv:0907.4673.

\bibitem{zyam-rpf-1}
N.~Sharma, J.~Mazer, M.~Stuart, and C.~Nattrass,
\newblock Phys. Rev. {\bf C93}, 044915 (2016), arXiv:1509.04732.

\bibitem{zyam-rpf-2}
C.~Nattrass, N.~Sharma, J.~Mazer, M.~Stuart, and A.~Bejnood,
\newblock Phys. Rev. {\bf C94}, 011901 (2016), arXiv:1606.00677.

\end{thebibliography}

\end{document}